\documentclass{nature}

\usepackage[pdftex]{graphicx}  
\usepackage{color, amsmath, amssymb, footmisc, authblk, url}
\usepackage[flushleft]{threeparttable}

\usepackage{afterpage,multirow,makecell,tabularx,caption,booktabs, array}
\usepackage{lineno}

\usepackage{soul}
\usepackage[normalem]{ulem}
\usepackage{pdfpages}

\title{Milky Way as a Neutrino Desert Revealed by IceCube Galactic Plane Observation}  
 
\author[1]{Ke Fang}
\author[2,3]{John S. Gallagher}
\author[1]{Francis Halzen}

\affil[1]{Department of Physics, Wisconsin IceCube Particle Astrophysics Center, University of Wisconsin, Madison, WI, 53706}

\affil[2]{
Department of Astronomy,  University of Wisconsin, Madison, WI, 53706
}
\affil[3]{Department of Physics and Astronomy, Macalester College, St. Paul., MN,  55105}

\begin{document}
\maketitle
\vspace {10pt}


\begin{abstract}

The Galactic diffuse emission (GDE) is formed when cosmic rays leave the sources where they were accelerated, diffusively propagate in the Galactic magnetic field, and interact with the interstellar medium and interstellar radiation field. GDE in $\gamma$-ray (GDE-$\gamma$) has been observed up to sub-PeV energies, though its origin may be explained by either cosmic-ray nuclei or electrons. We show that the $\gamma$-rays accompanying the high-energy neutrinos recently observed by the IceCube Observatory from the Galactic plane have a flux that is consistent with the GDE-$\gamma$ observed by the {\it Fermi}-LAT and Tibet AS$\gamma$ experiments around 1~TeV and 0.5~PeV, respectively. The consistency suggests that the diffuse $\gamma$-ray emission above $\sim$1~TeV could be dominated by hadronuclear interactions, though partial leptonic contribution cannot be excluded. Moreover, by comparing the fluxes of the Galactic and extragalactic diffuse emission backgrounds, we find that the neutrino luminosity of the Milky Way is one to two orders of magnitude lower than the average of distant galaxies. This implies that our Galaxy has not hosted the type of neutrino emitters that dominates the isotropic neutrino background at least in the past few tens of kiloyears. 
\end{abstract}

\maketitle


High-energy neutrinos have been observed from the Milky Way by the IceCube Observatory between 0.5~TeV and multi-PeV \cite{IceCubeGP, IceCube:2023hou}. 
The Galactic diffuse emission (GDE) in neutrinos (GDE-$\nu$) is identified at a $4.5\,\sigma$ significance using cascade data and templates describing the diffuse flux of photons. Unresolved individual sources also potentially contribute to the observed events. Below $\sim 10$~TeV, the GDE flux measured using the $\pi^0$ model, based on the MeV-to-GeV $\pi^0$ component measured by the {\it Fermi} Large Area Telescope (LAT) \cite{Fermi-LAT:2012edv}, and the CRINGE model \cite{2023ApJ...949...16S}, based on a global fit of cosmic rays, is higher than that using the KRA models  \cite{2015ApJ...815L..25G}, which implement radially-dependent cosmic-ray diffusion. Although the $\pi^0$ and CRINGE models are slightly favored by the data, the preference is not statistically significant \cite{IceCube:2023hou}.  

The Galactic diffuse emission in $\gamma$-rays (GDE-$\gamma$) has been measured by {\it Fermi}-LAT from 100~MeV to 1~TeV \cite{Fermi-LAT:2012edv}. Above $\sim$1~TeV, GDE-$\gamma$ has been observed by ground-based $\gamma$-ray experiments from the parts of the Galactic Plane that are accessible to the detectors \cite{HESS:2014ree, HAWC:2021bvb}. The GDE-$\gamma$ above 100~TeV has been detected by the Tibet AS$\gamma$  \cite{Tibet21}, HAWC \cite{HAWC:2021bvb}, and LHAASO-KM2A \cite{lhaasoGDE} observatories. Most of the photons observed by Tibet AS$\gamma$ with energy above 398~TeV do not point to known $\gamma$-ray sources, suggesting that the emission  could be diffuse in nature. On the other hand, LHAASO-KM2A finds a lower GDE intensity for a similar sky region when masking known and new sources detected by LHAASO \cite{lhaasoGDE}. This hints at the postulation that the Tibet AS$\gamma$ flux could partly be contributed by unresolved sources \cite{2022ApJ...928...19V}. 

GDE may come from protons and nuclei when they interact with gas in the interstellar medium (ISM). Diffuse $\gamma$-ray emission may also be produced by the inverse Compton radiation of relativistic electrons. The fraction of leptonic contribution to the GDE-$\gamma$ is still under debate. Significant inverse Compton radiation has been suggested to come from the Galactic bulge and inner Galaxy, contributing to both around a few GeV and above 10~TeV \cite{2017ApJ...846...67P, 2021arXiv211212745P}. A hardening in the diffuse $\gamma$-ray spectrum at 0.1-1~TeV,  sometimes referred to as the ``TeV excess", has been interpreted as a signature of unresolved TeV halos due to electrons trapped around middle-aged pulsars \cite{2018PhRvL.120l1101L} or pulsar winde nebulae \cite{2022CmPhy...5..161V}, or as a result of a progressive hardening of cosmic-ray nuclei spectra toward the Galactic center due to effects such as the anisotropic cosmic-ray transport \cite{2017JCAP...10..019C}.

\begin{figure} 
    \centering
   \includegraphics[width=0.75\textwidth]{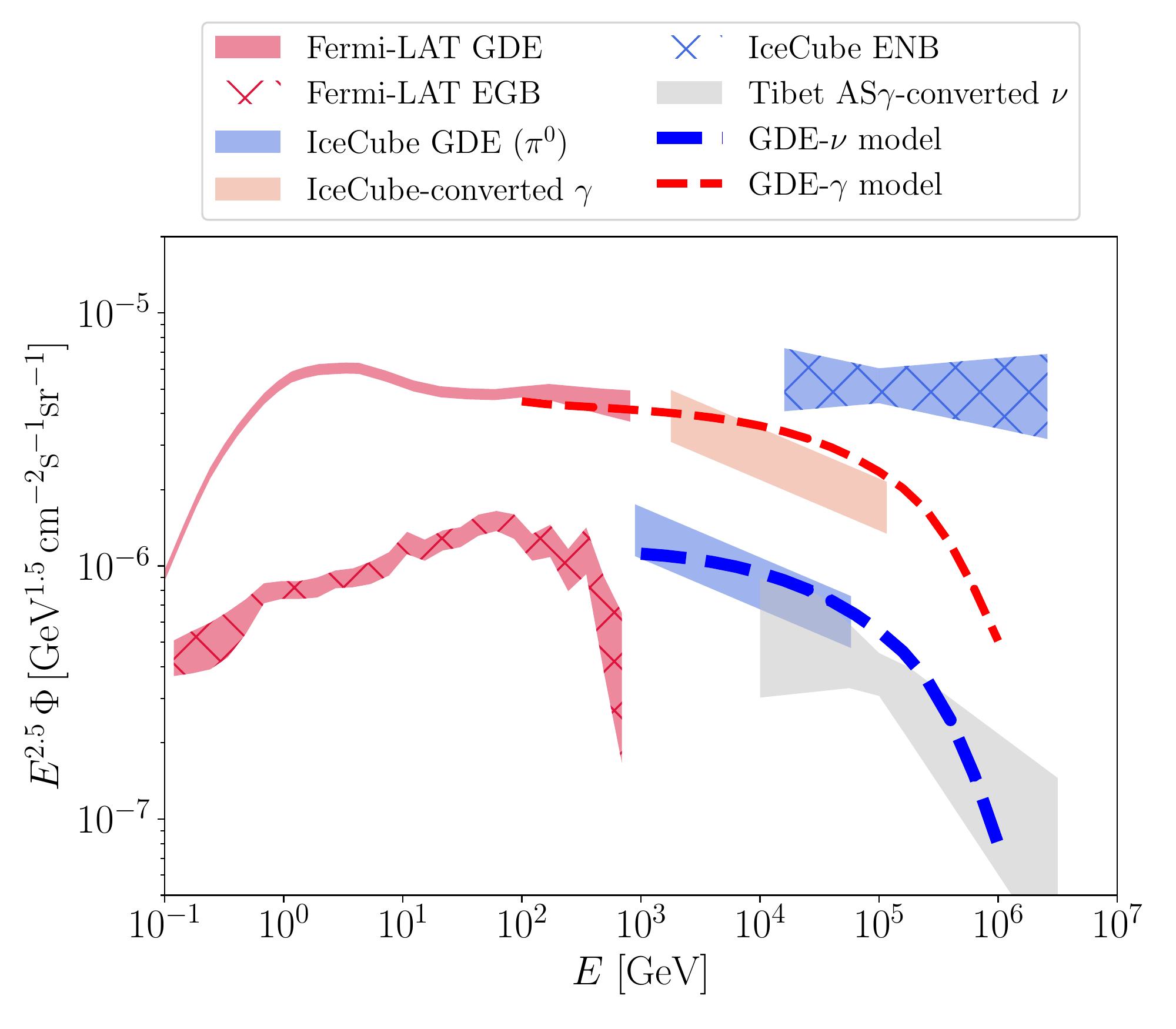}
    \caption{\label{fig:GalEx} {\bf All-sky-averaged intensities of the Galactic diffuse emission (GDE) and extragalactic background (EB) in $\gamma$-ray and neutrinos, scaled by $E^{2.5}$.}  Gamma-ray is indicated in red and neutrino (per-flavor flux including both neutrinos and antineutrinos) is in blue. Galactic components (unhatched regions) include the diffuse neutrino emission from the Galactic plane measured by IceCube using the $\pi^0$ template \cite{IceCubeGP} (blue shaded region indicating the $1\,\sigma$ uncertainties), the neutrino flux derived from the GDE-$\gamma$ measured by Tibet AS$\gamma$ \cite{Tibet21, FangMurase21} (silver shaded region), and the Galactic interstellar emission model of {\it Fermi}-LAT \cite{4FGL} (red shaded region). The dashed curves present a numerical simulation of the diffuse emission that accounts for the spatial distribution of sources and gas in the Milky Way. Extragalactic components (hatched regions) include the isotropic diffuse neutrino background measured by IceCube \cite{cascade6yr} (blue hatched region) and the extragalactic $\gamma$-ray background measured by {\it Fermi}-LAT \cite{FermiIGRB} (red hatched region).  
    } 
\end{figure}

Cosmic-ray protons and nuclei interact with the ISM gas and produce neutrinos and $\gamma$-rays simultaneously via $\pi^\pm\rightarrow e^\pm\nu_e (\bar{\nu}_e)\bar{\nu}_\mu \nu_\mu$ and $\pi^0\rightarrow 2\gamma$. The flux of the $\gamma$-rays that accompany the IceCube Galactic diffuse neutrinos can be estimated  by $ E_\gamma^2 ({dN_{\rm inj}}/{dE_\gamma}) \approx  {2}/{3} E_\nu^2  ({dN}/{dE_\nu})$, with $  {E_\gamma \approx 2\,{E_\nu}}$ \cite{Ahlers:2013xia}. Here ${dN_{\rm inj}}/{dE_\gamma}$ denotes the injected $\gamma$-ray spectrum, which could be different from the observed $\gamma$-ray spectrum due to the pair production of $\gamma$-rays on low-energy photons. Photons above $\sim 100$~TeV may be absorbed by the interstellar radiation field, while the attenuation effect is negligible for lower-energy photons \cite{2018PhRvD..98d3003L, FangMurase21}. As shown in Figure~\ref{fig:GalEx}, the $\gamma$-ray flux derived from the IceCube measurement  is comparable to the {\it Fermi}-LAT Galactic interstellar emission model around 1~TeV (note that the terms ``interstellar $\gamma$-ray emission" and ``Galactic diffuse emission" both refer to the diffuse emission made by energetic cosmic rays interacting with interstellar nucleons and photons \cite{2016ApJS..223...26A}). The model is obtained by fitting various templates of the diffuse $\gamma$-ray emission and models of resolved and unresolved sources to the {\it Fermi}-LAT data between 100~MeV and 1~TeV \cite{4FGL}. 
The systematic error in the effective area of {\it Fermi}-LAT Pass~8 data is estimated to be 5\% between 0.1 and 100~GeV, and 15\% at 1~TeV with a linear interpolation in logarithm of energy between 100~GeV and 1~TeV. 
We used the systematic error to estimate the uncertainty of the Galactic interstellar emission spectrum.  The actual measurement error may also arise from the model uncertainties and the separation of the isotropic emission \cite{2016ApJS..223...26A}, and thus  be higher than the systematics of the detector.

The flux of GDE-$\nu$ observed by IceCube is also consistent with that of the diffuse neutrinos expected to accompany the sub-PeV diffuse $\gamma$-rays observed by Tibet AS$\gamma$. The shaded silver region in Figure~\ref{fig:GalEx}, from reference \cite{FangMurase21}, is an estimation of the Galactic plane emission derived from the Tibet AS$\gamma$ measurements in two sky regions, namely, {\it region A}: $25^\circ < l<100^\circ$ and {\it region B:} $50^\circ < l<200^\circ$, both with  $|b| < 5^\circ$.  The width of the band accounts for the uncertainties due to the spectra and spatial distribution of cosmic rays, gas density, and infrared emission of the ISM. The IceCube observation between 10 and $\sim$60~TeV agrees with this Tibet-converted neutrino flux within the uncertainties.

The consistency in the GDE measurements by {\it Fermi}-LAT, Tibet AS$\gamma$, and IceCube at various energies suggests that the GDE-$\gamma$ could be dominantly produced by hadronic interaction above $\sim$1~TeV. However, given the uncertainty of the GDE-$\nu$ flux associated with the analysis templates and the potential contribution from unresolved neutrino sources, leptonic processes may still play a role in particular between 0.1 and a few TeV.

Below we consider a GDE model under the assumption that the IceCube flux based on the $\pi^0$ template represents the diffuse neutrino flux and the source contribution is negligible. The modeling of diffuse neutrino and $\gamma$-ray emission is impacted by several poorly known factors, including 1) cosmic-ray spectra above the rigidity $\sim$10 TV observed at the Earth, 2) the difference in the cosmic-ray density at a location $\vec{x}$ in the Galaxy and that at the observer point, $n(\vec{x}, E) / n(r_\odot, E)$, which is determined by  the source distribution in the Galaxy, timescales of the sources, and the particle diffusion in the Galactic magnetic field, and 3) the density profile of the neutral, ionized, and dark gas. The effects of these factors could be coupled and are not constrained by current observations, as also noted by e.g., references \cite{2021arXiv211212745P,2017JCAP...02..015E, 2022arXiv220315759D, 2023ApJ...949...16S}. 

To limit the degrees of freedom of our model, as in reference~\cite{2016JCAP...11..004P, 2018PhRvD..98d3003L, 2019JCAP...12..050C} we consider a simplified model where the spatial and spectral components of the nucleon flux are assumed to be independent, $\Phi(\vec{x}, E) = \Phi(\vec{x}_\odot, E)\, \left[n(\vec{x}) / n(\vec{x}_\odot)\right]$. We obtain the ratio of cosmic-ray density at position $\vec{x}$ to that at the solar neighborhood, $n(\vec{x}) / n(\vec{x}_\odot)$, using a numerical simulation that propagates cosmic rays  from synthetic sources in the Galactic magnetic field. We fix the spectra of protons and helium nuclei at the solar neighborhood, $\Phi(\vec{x}_\odot, E)$, to the best-fit model obtained by fitting to the cosmic-ray measurements between $\sim 10$~GeV and $\sim$10~PeV. More details about the simulation and the calculation of the intensity of neutrino and $\gamma$-ray emission are explained in Supplemental Material's Section~1 and 2, respectively.

The dashed curves in Figure~\ref{fig:GalEx} from our model show that the hadronic interaction may simultaneously explain the observed $\gamma$-ray spectra between $100$ GeV to 100~TeV and the diffuse neutrino flux of the Galactic Plane measured using the $\pi^0$ template. More complicated models including multiple components of cosmic-ray sources and diffusion regions \cite{2017JCAP...02..015E, 2023ApJ...949...16S} may provide better fits to the data. 
Future measurements of longitudinal and latitudinal profiles of neutrino emission above 1~TeV, identification of individual Galactic neutrino sources, and observation of GDE-$\gamma$ from the Southern sky at TeV-PeV energies are needed to break down the degeneracy of the model parameters.



The GDE flux reflects the emissivity of our own Galaxy in high-energy neutrinos while the extragalactic background (EB) reveals the contribution of powerful sources in distant galaxies. Had the local and distant sources been similarly luminous, the GDE would be brighter than the EB due to geometry. 
Figure~\ref{fig:GalEx} contrasts the intensities of the GDE and EB in $\gamma$-ray and neutrinos. Notably, the all-sky averaged GDE-$\gamma$ is brighter than the extragalactic $\gamma$-ray background (EGB) between 1~GeV and 1~TeV, whereas the GDE-$\nu$ is fainter than the extragalactic neutrino background (ENB) between 1 and 100~TeV. The fact suggests that the Milky Way in its current state is not a typical source of high-energy neutrinos. 

The integrated differential flux at the neutrino energy $E_\nu$ observed today from galaxies extending to cosmic ``high noon"  ($z\sim2-3$) where the star formation rate peaks can be calculated as 
\begin{eqnarray}\label{eqn:Phi_EG}
   E_\nu^2 \Phi_\nu^{\rm EG} (E_\nu) &=& \frac{c}{4\pi} \int_{0}^{z_{\rm max}} dz  \Big|\frac{dt}{dz}\Big|   \frac{1}{1+z} 
  \int_{M_{\rm min}}^{M_{\rm max}} dM \frac{dn}{dM} (M, z) 
  L_\nu^{\rm EG}(E'_\nu, M, z),  
\end{eqnarray}
here $E_\nu'=E_\nu(1+z)$ is the energy of a neutrino at redshift $z$, $|dt/dz| = (H_0 (1+z) \sqrt{\Omega_M (1+z)^3 + \Omega_\Lambda})^{-1}$ with $H_0=67.4\,\rm km\,s^{-1}\,Mpc^{-1}$, $\Omega_M=0.315$, and assuming a flat universe \cite{2020A&A...641A...6P}, and $L_\nu^{\rm EG} (E'_\nu, M, z)\equiv E_\nu^{'2} d\dot{N}_\nu/dE'_\nu (M, z)$ is the neutrino luminosity of an external galaxy.  The number density $n$ of galaxies with stellar mass $M$ at redshift z is $dn/dM (M, z)$ given by the Schechter \cite{1976ApJ...203..297S} function, 
${dn}/{d\log M}= \phi^*\,\ln (10) (10^{\log M - \log M^*})^{\alpha + 1}   \exp (-10^{\log M - \log M^*})$. The normalization $\phi^*$, slope $\alpha$, and characteristic mass $M^*$ are found by fitting the mass function to galaxy distribution at different redshift bins up to $z\sim 3$ \cite{2006A&A...445..805F, 2014ApJ...783...85T, 2016ApJ...830...83C}.

We assume that the neutrino emissivity of a galaxy, including contributions from both the individual sources hosted by the galaxy and the galactic diffuse emission, is related to the stellar mass and redshift independently:
\begin{eqnarray} \label{eqn:L_nu_EG}
    L_\nu^{\rm EG}(E'_\nu, M, z) &=& L_\nu^{\rm EG}(E_\nu, M_{\rm MW},  0) 
     \left(\frac{M}{M_{\rm MW}}\right)^{\beta}g(z)(1+z)^{2-s}.
\end{eqnarray}In the above expression, we have parameterized the dependence of the stellar mass as a power-law with an index $\beta$ depending on source models. In an extreme scenario where all galaxies are equally luminous, $\beta = 0$. In a more realistic scenario where the neutrino luminosity scales to the optical, near infrared, or X-ray luminosity of galaxies, one would expect $\beta\sim 1$ \cite{2003ApJS..149..289B, 2004MNRAS.349..146G}. The $\gamma$-ray luminosities of the star-forming galaxies detected by {\it Fermi}-LAT present a relation of $L_\gamma\propto L_{\rm IR}^{1.35}$ \cite{2020A&A...641A.147K}. If the neutrino luminosity is proportional to the bolometric luminosity of the AGN, then $\beta$ may reach as high as $1.47$ \cite{2020ApJ...889...32S}. 
Without specifying the neutrino source types, below we float $\beta$ from 0 to 2.0. 

The function $g(z)$ in equation~\ref{eqn:L_nu_EG} describes the source evolution over redshift. In a uniform evolution scenario, $g(z) = 1$. If the sources follow a star-formation history, $g(z)$ can be modelled as $g(z)\propto (1+z)^{3.4}$ at $z<1$, $(1+z)^{-0.3}$ at $1<z<4$, and $ (1+z)^{-3.5}$ at $z>4$ \cite{2006ApJ...651..142H}. 
A star-formation model with higher redshift contributions enhances the integrated flux by a factor of order unity. Our calculation considers both the uniform and star-formation models. 

The last term in equation~\ref{eqn:L_nu_EG} arises from the fact that a neutrino observed at $E_\nu$ today was at the energy $E_\nu (1+z)$ at the source. So $E_\nu^{'2}dN/dE'_\nu \propto (1+z) ^{2-s}$ when assuming that the neutrino spectrum follows a non-broken power-law $dN_\nu/dE_\nu \propto E_\nu^{-s}$. Measurements of the diffuse isotropic neutrino flux find the index $s\sim 2.5$ between $\sim 1$~TeV and a few PeV \cite{HESE75yr, cascade6yr, numu95yr}. 
Without loss of generality, we take $s = 2.0-3.0$.   





As explained in the Methods, the all-sky-averaged intensity of the GDE observed at the solar neighborhood may be related to the total neutrino power of the Milky Way through $E_\nu^2\Phi_\nu^{\rm MW} (E_\nu) = F_\epsilon (E_\nu) ({3}/{4\pi)} \left({L_\nu^{\rm MW}(E_\nu)}/ {4\pi r_\odot^2}\right)$ with $r_\odot\approx 8.5\,\rm kpc$ being our distance to the Galactic center and $F_\epsilon$ a geometry factor of the order unity that accounts for the profiles of gas and sources.

Equations~\ref{eqn:Phi_EG} and \ref{eqn:L_nu_EG} may also apply to $\gamma$-rays when the attenuation due to $\gamma\gamma$ pair production with the cosmic microwave background and the interstellar radiation field is negligible, which is the case below $\sim$100~GeV. Note that equation~\ref{eqn:L_nu_EG} makes no assumption on the production mechanism of the $\gamma$-rays and thus holds true regardless of their hadronic or leptonic origin. When deriving the total luminosity of the Milky Way using the GDE, we have assumed that the contribution of resolved and unresolved sources in the Galaxy is negligible. This is consistent with the observation of {\it Fermi}-LAT up to $\sim 100$~GeV \cite{Fermi-LAT:2012edv, 2016ApJS..223...26A}.

The ratio of the luminosities of an external, Milky Way-like galaxy to the Milky Way, $L^{\rm EG} / L^{\rm MW}\equiv L^{\rm EG}(M_{\rm MW}, 0) / L^{\rm MW}$ in neutrinos can be estimated as
\begin{eqnarray}\label{eq:ratioEstimate}
   \frac{L_\nu^{\rm EG}}{L_\nu^{\rm MW}} &\approx& \frac{\Phi_\nu^{\rm EG}}{\Phi_\nu^{\rm MW}}  \frac{3 F_\epsilon}{4\pi r_\odot^2  n_0 c t_H  \xi_z} = 
   120 \,\left(\frac{{\Phi_\nu^{\rm EG}}/{\Phi_\nu^{\rm MW}}}{5}\right) \left(\frac{n_0}{0.01\,\rm Mpc^{-3}}\right)^{-1}\left(\frac{\xi_z}{3}\right)^{-1}\left(\frac{F_\epsilon}{1}\right),
\end{eqnarray}
where $n_0$ is the local density of galaxies with a similar stellar mass as the Milky Way and the quantity $\xi_z$  accounts for the evolution of the source emissivity over cosmic time \cite{1998PhRvD..59b3002W}. $\xi_z$ is determined by $g(z)$ and varies from $\sim 0.5$ for uniform evolution to $\sim 3$ for star-formation evolution. While the analytical expression in equation~\ref{eq:ratioEstimate} demonstrates the dependence on various parameters, the extragalactic-Galactic ratio presented below is numerically computed using equations~\ref{eqn:Phi_EG}, \ref{eqn:L_nu_EG},  \ref{eqn:phinu_GC} and \ref{eqn:emissivity}. 
 
\begin{figure} 
    \centering
   \includegraphics[width=0.75\textwidth]{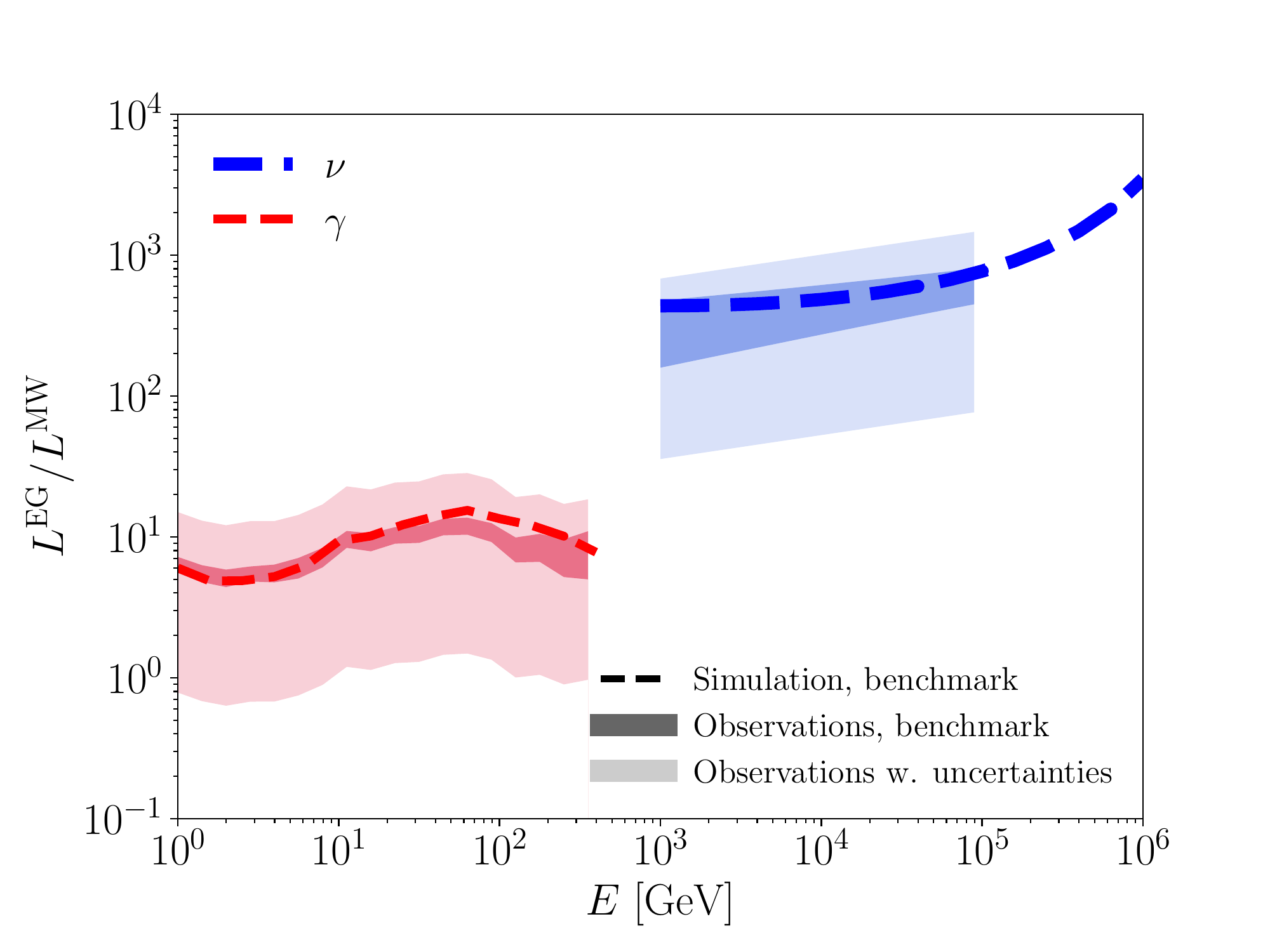}
    \caption{\label{fig:GalExRatio} {\bf Derived ratio of the average luminosity of an external, Milky Way-like galaxy and the Galactic luminosity in GeV-TeV $\gamma$-ray and TeV-PeV neutrinos.} The ratio is calculated using equations~\ref{eqn:Phi_EG},  \ref{eqn:L_nu_EG}, \ref{eqn:phinu_GC} and \ref{eqn:emissivity} with the {\it Fermi}-LAT and IceCube measurements of the GDE. The dark shaded regions adopt the benchmark extragalactic model parameters and  assume a uniform emissivity within the Galactic disk. Their widths correspond to the uncertainties in the GDE and EB observations. The dashed curves also use the benchmark extragalactic model but  a more realistic Galactic model that takes into account the spatial distribution of sources and gas in the Milky Way. The light shaded region further accounts for the uncertainties in the parameters of the extragalactic model.  
    } 
\end{figure}

Figure~\ref{fig:GalExRatio} presents $L^{\rm EG} / L^{\rm MW}$  derived from TeV-PeV neutrino and GeV-TeV $\gamma$-ray observations. For the extragalactic diffuse background, we take $\beta = 1$, $s = 2.5$, and $g(z)$ following the star-formation history as the benchmark model. 
The dark shaded regions show the results obtained with the extragalactic benchmark model and by assuming a uniform gamma-ray and neutrino emissivity ($\epsilon_{\gamma,\nu}$ in equation~\ref{eqn:emissivity}) inside the Galactic Plane. 
The width of the dark shaded region are propagated errors from the observational uncertainties of the ENB measured with the IceCube 6-year cascade events \cite{cascade6yr}, the GDE-$\nu$ flux found by IceCube with the $\pi^0$ diffusion template \cite{IceCubeGP}, and the EGB measured by {\it Fermi}-LAT assuming foreground model A \cite{FermiIGRB}. 
For comparison, the dashed curve shows the ratio derived with the same extragalactic model parameters but with a more realistic GDE model from the simulation. 
The light shaded region additional accounts for the uncertainties in the parameters $\beta$, $s$, and $g(z)$. 
In all cases, we find that $L_\gamma^{\rm EG}$ is not significantly different from $L_\gamma^{\rm MW}$ but $L_\nu^{\rm EG}/L_\nu^{\rm MW}\gg 1$. In other words, the Milky Way at the present time is an atypical neutrino emitter. The IceCube observation of the GDE confines this ratio to $\sim 30-10^3$ depending on the neutrino energy. Our finding also suggests that GeV $\gamma$-rays are suppressed relative to the neutrinos observed by IceCube. GeV $\gamma$-rays are either barely produced in the process where TeV neutrinos are generated or, the accompanying gamma rays are attenuated by the radiation field at  the neutrino production site. 





Our result suggests that the Galaxy has not hosted the type of emitters that dominate the ENB in the past $D / c \sim 26\,{\rm kyr} \,(D / 8\,\rm kpc)$, which is the time taken by a neutrino to travel from a Galactic source at a kpc-scale distance $D$ to the Earth. Cosmic rays at TeV-PeV energy are confined by the Galactic magnetic field for million-year durations \cite{2007ARNPS..57..285S}. Had any major cosmic-ray sources injected protons into the ISM within that time period, the diffuse neutrino flux of the Galactic plane would be higher and the gap in the neutrino luminosity of our Galaxy and an external Milky Way-like galaxy would be smaller. 



There is compelling evidence for a highly energetic Seyfert explosion from the supermassive black hole at the Galactic center a few million years in the past. Among that, the clearest indications are the Fermi and eROSITA and bubbles \cite{Fermi-LAT:2014sfa, Predehl:2020kyq}. The time to the last burst/flare is constrained to 2-10 Myr by both the the mechanical timescales needed to explain the morphology and multi-wavelength spectra of the observed bubbles and haze \cite{Yang:2022jck} and kinematic studies of halo gas \cite{2015ApJ...799L...7F, 2016ApJ...829....9M, 2017ApJ...834..191B}. 
In addition, elevated ionizing radiation along the Magellanic Stream \cite{2019ApJ...886...45B} independently constrains this nuclear activity. These timescales are consistent with the scenario where most PeV protons from the last jet activity have already left the Galaxy today and the Milky Way is no longer an active neutrino emitter. Other sources or mechanisms that are not present or extremely rare in the Milky Way over the past tens of thousands of years, such as tidal disruption events, could also contribute to the extragalactic neutrino background. These sources are likely $\gamma$-ray-obscured as suggested by both the observations of the isotropic neutrino flux \cite{2016PhRvL.116g1101M,2022ApJ...933..190F} and individual neutrino sources \cite{2022Sci...378..538I}.  


\begin{addendum}
\item[Methods]{$ $}  

{\bf Neutrino Emissivity and Intensity of the Galaxy} 

The all-sky-averaged intensity of the Galactic Plane observed at the solar neighborhood is
\begin{eqnarray}\label{eqn:phinu_GC} 
   E_\nu^2\Phi_\nu^{\rm MW} (E_\nu) &=& \frac{1}{4\pi}\int_{-\pi/2}^{\pi/2} \cos(b) db \int_0^{2\pi} dl  I_\nu(l, b, E_\nu)  
\end{eqnarray}
where $I_\nu(l,b, E_\nu)=(1/4\pi)\int_0^{\infty}  ds\, \epsilon_\nu (l, b, r, E_\nu)$ is the intensity from the direction with Galactic longitude and latitude $(l,b)$ along a line of sight $s$ of the observer, and $\epsilon_\nu (l, b, r, E_\nu)$ is the production rate of neutrinos per unit volume in units of $\rm eV\,s^{-1}\,cm^{-3}$ at a distance $r$ from the Galactic center, which is related to the total neutrino power emitted by the Milky Way, $L_\nu^{\rm MV}$, through
\begin{equation}\label{eqn:emissivity}
    L_\nu^{\rm MW} (E_\nu)= \int_{-\pi/2}^{\pi/2} \cos(b) db \int_0^{2\pi} dl \int_0^{\infty} r^2 dr\, \epsilon_\nu (l, b, r, E_\nu). 
\end{equation}
Unresolved point-like and extended sources may contribute to both the Tibet AS$\gamma$ and IceCube observations \cite{IceCubeGP, Tibet21}. Equations~\ref{eqn:phinu_GC} and \ref{eqn:emissivity} still apply in the presence of individual sources. Like equations~\ref{eqn:Phi_EG} and \ref{eqn:L_nu_EG}, these two equations may also apply to $\gamma$-rays when the $\gamma$-ray absorption by the interstellar radiation field is negligible.

The intensity of the Galactic Plane depends on the emission profile. We can rewrite equations~ \ref{eqn:phinu_GC} and \ref{eqn:emissivity} as $E_\nu^2\Phi_\nu^{\rm MV} (E_\nu) = F_\epsilon (E_\nu) ({3}/{4\pi)} \left({L_\nu^{\rm MW}(E_\nu)}/ {4\pi r_\odot^2}\right)$ with $F_\epsilon$ denoting a geometry factor of the order unity that accounts for the profiles of gas and sources. It essentially says that the total flux of GDE observed at the solar neighborhood is, to the first order, comparable to the flux from a point source at the Galactic center that carries the power of the entire Galaxy. When assuming that $\epsilon_\nu$ is independent on $E_\nu$ and uniform in the Galactic disk as in the leaky box model, with the disk defined as $R_{\rm d} < 15$~kpc and $z_{\rm d}<0.2$~kpc, we obtain $F_\epsilon = 1.16$. Alternatively, we find $F_\epsilon = 0.97$ when assuming that $\epsilon_\nu$ follows the spatial distribution of supernova remnants \cite{FangMurase21}.

\item[Data Availability]
Source data for Figure~1 are attached. Any additional data are available from the corresponding author upon reasonable request.

\item[Code Availability]
The calculation used publicly available software packages including HERMES (\url{https://github.com/cosmicrays/hermes}) and CRPropa (\url{https://github.com/CRPropa}).

\item[Acknowledgements]
The work of K.F and F.H is supported by the Office of the Vice Chancellor for Research and Graduate Education at the University of Wisconsin-Madison with funding from the Wisconsin Alumni Research Foundation. K.F. acknowledges support from National Science Foundation (PHY-2110821, PHY-2238916) and NASA (NMH211ZDA001N-Fermi). This work was supported by a grant from the Simons Foundation (00001470, KF). J.S.G. thanks the University of Wisconsin College of Letters and Science for partial support of his IceCube-related research. The research of F.H was also supported in part by the U.S. National Science Foundation under grants~PHY-2209445 and OPP-2042807.

\item[Author Contributions Statement]
K.F. carried out the simulations and analyses and prepared the manuscript. All authors participated in the interpretation of the results and edited the manuscript.

\item[Competing Interests Statement]
The authors declare no competing interests.

\end{addendum}

\clearpage




\newcommand{\actaa}{Acta Astron.}   
\newcommand{\araa}{Annu. Rev. Astron. Astrophys.}   
\newcommand{\areps}{Annu. Rev. Earth Planet. Sci.} 
\newcommand{\aar}{Astron. Astrophys. Rev.} 
\newcommand{\ab}{Astrobiol.}    
\newcommand{\aj}{Astron. J.}   
\newcommand{\ac}{Astron. Comput.} 
\newcommand{\apart}{Astropart. Phys.} 
\newcommand{\apj}{Astrophys. J.}   
\newcommand{\apjl}{Astrophys. J. Lett.}   
\newcommand{\apjs}{Astrophys. J. Suppl. Ser.}   
\newcommand{\ao}{Appl. Opt.}   
\newcommand{\apss}{Astrophys. Space Sci.}   
\newcommand{\aap}{Astron. Astrophys.}   
\newcommand{\aapr}{Astron. Astrophys. Rev.}   
\newcommand{\aaps}{Astron. Astrophys. Suppl.}   
\newcommand{\baas}{Bull. Am. Astron. Soc.}   
\newcommand{\caa}{Chin. Astron. Astrophys.}   
\newcommand{\cjaa}{Chin. J. Astron. Astrophys.}   
\newcommand{\cqg}{Class. Quantum Gravity}    
\newcommand{\epsl}{Earth Planet. Sci. Lett.}    
\newcommand{\frass}{Front. Astron. Space Sci.}    
\newcommand{\gal}{Galaxies}    
\newcommand{\gca}{Geochim. Cosmochim. Acta}   
\newcommand{\grl}{Geophys. Res. Lett.}   
\newcommand{\icarus}{Icarus}   
\newcommand{\jatis}{J. Astron. Telesc. Instrum. Syst.}  
\newcommand{\jcap}{J. Cosmol. Astropart. Phys.}   
\newcommand{\jgr}{J. Geophys. Res.}   
\newcommand{\jgrp}{J. Geophys. Res.: Planets}    
\newcommand{\jqsrt}{J. Quant. Spectrosc. Radiat. Transf.} 
\newcommand{\lrca}{Living Rev. Comput. Astrophys.}    
\newcommand{\lrr}{Living Rev. Relativ.}    
\newcommand{\lrsp}{Living Rev. Sol. Phys.}    
\newcommand{\memsai}{Mem. Soc. Astron. Italiana}   
\newcommand{\mnras}{Mon. Not. R. Astron. Soc.}   
\newcommand{\nat}{Nature} 
\newcommand{\nastro}{Nat. Astron.} 
\newcommand{\ncomms}{Nat. Commun.} 
\newcommand{\nphys}{Nat. Phys.} 
\newcommand{\na}{New Astron.}   
\newcommand{\nar}{New Astron. Rev.}   
\newcommand{\physrep}{Phys. Rep.}   
\newcommand{\pra}{Phys. Rev. A}   
\newcommand{\prb}{Phys. Rev. B}   
\newcommand{\prc}{Phys. Rev. C}   
\newcommand{\prd}{Phys. Rev. D}   
\newcommand{\pre}{Phys. Rev. E}   
\newcommand{\prl}{Phys. Rev. Lett.}   
\newcommand{\psj}{Planet. Sci. J.}   
\newcommand{\planss}{Planet. Space Sci.}   
\newcommand{\pnas}{Proc. Natl Acad. Sci. USA}   
\newcommand{\procspie}{Proc. SPIE}   
\newcommand{\pasa}{Publ. Astron. Soc. Aust.}   
\newcommand{\pasj}{Publ. Astron. Soc. Jpn}   
\newcommand{\pasp}{Publ. Astron. Soc. Pac.}   
\newcommand{\raa}{Res. Astron. Astrophys.} 
\newcommand{\rmxaa}{Rev. Mexicana Astron. Astrofis.}   
\newcommand{\sci}{Science} 
\newcommand{\sciadv}{Sci. Adv.} 
\newcommand{\solphys}{Sol. Phys.}   
\newcommand{\sovast}{Soviet Astron.}   
\newcommand{\ssr}{Space Sci. Rev.}   
\newcommand{\uni}{Universe} 

\newcommand\mathplus{+}


\begin{thebibliography}{10}
\expandafter\ifx\csname url\endcsname\relax
  \def\url#1{\texttt{#1}}\fi
\expandafter\ifx\csname urlprefix\endcsname\relax\def\urlprefix{URL }\fi
\providecommand{\bibinfo}[2]{#2}
\providecommand{\eprint}[2][]{\url{#2}}

\bibitem{IceCubeGP}
\bibinfo{author}{{IceCube Collaboration}}.
\newblock \bibinfo{title}{Observation of high-energy neutrinos from the
  galactic plane}.
\newblock \emph{\bibinfo{journal}{Science}} \textbf{\bibinfo{volume}{380}},
  \bibinfo{pages}{1338--1343} (\bibinfo{year}{2023}).
\newblock

\bibitem{IceCube:2023hou}
\bibinfo{author}{Fuerst, P.~M.} \emph{et~al.}
\newblock \bibinfo{title}{{Galactic and Extragalactic Analysis of the
  Astrophysical Muon Neutrino Flux with 12.3 years of IceCube Track Data}}.
\newblock \emph{\bibinfo{journal}{PoS}} \textbf{\bibinfo{volume}{ICRC2023}},
  \bibinfo{pages}{1046} (\bibinfo{year}{2023}).

\bibitem{Fermi-LAT:2012edv}
\bibinfo{author}{Ackermann, M.} \emph{et~al.}
\newblock \bibinfo{title}{{Fermi-LAT Observations of the Diffuse Gamma-Ray
  Emission: Implications for Cosmic Rays and the Interstellar Medium}}.
\newblock \emph{\bibinfo{journal}{Astrophys. J.}}
  \textbf{\bibinfo{volume}{750}}, \bibinfo{pages}{3} (\bibinfo{year}{2012}).

\bibitem{2023ApJ...949...16S}
\bibinfo{author}{{Schwefer}, G.}, \bibinfo{author}{{Mertsch}, P.} \&
  \bibinfo{author}{{Wiebusch}, C.}
\newblock \bibinfo{title}{{Diffuse Emission of Galactic High-energy Neutrinos
  from a Global Fit of Cosmic Rays}}.
\newblock \emph{\bibinfo{journal}{\apj}} \textbf{\bibinfo{volume}{949}},
  \bibinfo{pages}{16} (\bibinfo{year}{2023}).

\bibitem{2015ApJ...815L..25G}
\bibinfo{author}{{Gaggero}, D.}, \bibinfo{author}{{Grasso}, D.},
  \bibinfo{author}{{Marinelli}, A.}, \bibinfo{author}{{Urbano}, A.} \&
  \bibinfo{author}{{Valli}, M.}
\newblock \bibinfo{title}{{The Gamma-Ray and Neutrino Sky: A Consistent Picture
  of Fermi-LAT, Milagro, and IceCube Results}}.
\newblock \emph{\bibinfo{journal}{\apjl}} \textbf{\bibinfo{volume}{815}},
  \bibinfo{pages}{L25} (\bibinfo{year}{2015}).

\bibitem{HESS:2014ree}
\bibinfo{author}{Abramowski, A.} \emph{et~al.}
\newblock \bibinfo{title}{{Diffuse Galactic gamma-ray emission with H.E.S.S}}.
\newblock \emph{\bibinfo{journal}{Phys. Rev. D}} \textbf{\bibinfo{volume}{90}},
  \bibinfo{pages}{122007} (\bibinfo{year}{2014}).

\bibitem{HAWC:2021bvb}
\bibinfo{author}{Abeysekara, A.~U.} \emph{et~al.}
\newblock \bibinfo{title}{{Galactic Gamma-Ray Diffuse Emission at TeV energies
  with HAWC Data}}.
\newblock \emph{\bibinfo{journal}{PoS}} \textbf{\bibinfo{volume}{ICRC2021}},
  \bibinfo{pages}{835} (\bibinfo{year}{2021}).

\bibitem{Tibet21}
\bibinfo{author}{{Tibet AS${\gamma}$ Collaboration}} \emph{et~al.}
\newblock \bibinfo{title}{{First Detection of sub-PeV Diffuse Gamma Rays from
  the Galactic Disk: Evidence for Ubiquitous Galactic Cosmic Rays beyond PeV
  Energies}}.
\newblock \emph{\bibinfo{journal}{\prl}} \textbf{\bibinfo{volume}{126}},
  \bibinfo{pages}{141101} (\bibinfo{year}{2021}).

\bibitem{lhaasoGDE}
\bibinfo{author}{{Cao}, Z.} \emph{et~al.}
\newblock \bibinfo{title}{{Measurement of ultra-high-energy diffuse gamma-ray
  emission of the Galactic plane from 10 TeV to 1 PeV with LHAASO-KM2A}}.
\newblock \emph{\bibinfo{journal}{arXiv e-prints}}
  \bibinfo{pages}{arXiv:2305.05372} (\bibinfo{year}{2023}).

\bibitem{2022ApJ...928...19V}
\bibinfo{author}{{Vecchiotti}, V.}, \bibinfo{author}{{Zuccarini}, F.},
  \bibinfo{author}{{Villante}, F.~L.} \& \bibinfo{author}{{Pagliaroli}, G.}
\newblock \bibinfo{title}{{Unresolved Sources Naturally Contribute to PeV
  Gamma-Ray Diffuse Emission Observed by Tibet AS{\ensuremath{\gamma}}}}.
\newblock \emph{\bibinfo{journal}{\apj}} \textbf{\bibinfo{volume}{928}},
  \bibinfo{pages}{19} (\bibinfo{year}{2022}).

\bibitem{2017ApJ...846...67P}
\bibinfo{author}{{Porter}, T.~A.}, \bibinfo{author}{{J{\'o}hannesson}, G.} \&
  \bibinfo{author}{{Moskalenko}, I.~V.}
\newblock \bibinfo{title}{{High-energy Gamma Rays from the Milky Way:
  Three-dimensional Spatial Models for the Cosmic-Ray and Radiation Field
  Densities in the Interstellar Medium}}.
\newblock \emph{\bibinfo{journal}{\apj}} \textbf{\bibinfo{volume}{846}},
  \bibinfo{pages}{67} (\bibinfo{year}{2017}).

\bibitem{2021arXiv211212745P}
\bibinfo{author}{{Porter}, T.~A.}, \bibinfo{author}{{J{\'o}hannesson}, G.} \&
  \bibinfo{author}{{Moskalenko}, I.~V.}
\newblock \bibinfo{title}{{The GALPROP Cosmic-ray Propagation and Nonthermal
  Emissions Framework: Release v57}}.
\newblock \emph{\bibinfo{journal}{\apjs}} \textbf{\bibinfo{volume}{262}},
  \bibinfo{pages}{30} (\bibinfo{year}{2022}).

\bibitem{2018PhRvL.120l1101L}
\bibinfo{author}{{Linden}, T.} \& \bibinfo{author}{{Buckman}, B.~J.}
\newblock \bibinfo{title}{{Pulsar TeV Halos Explain the Diffuse TeV Excess
  Observed by Milagro}}.
\newblock \emph{\bibinfo{journal}{\prl}} \textbf{\bibinfo{volume}{120}},
  \bibinfo{pages}{121101} (\bibinfo{year}{2018}).

\bibitem{2022CmPhy...5..161V}
\bibinfo{author}{{Vecchiotti}, V.}, \bibinfo{author}{{Pagliaroli}, G.} \&
  \bibinfo{author}{{Villante}, F.~L.}
\newblock \bibinfo{title}{{The contribution of Galactic TeV pulsar wind nebulae
  to Fermi large area telescope diffuse emission}}.
\newblock \emph{\bibinfo{journal}{Communications Physics}}
  \textbf{\bibinfo{volume}{5}}, \bibinfo{pages}{161} (\bibinfo{year}{2022}).

\bibitem{2017JCAP...10..019C}
\bibinfo{author}{{Cerri}, S.~S.}, \bibinfo{author}{{Gaggero}, D.},
  \bibinfo{author}{{Vittino}, A.}, \bibinfo{author}{{Evoli}, C.} \&
  \bibinfo{author}{{Grasso}, D.}
\newblock \bibinfo{title}{{A signature of anisotropic cosmic-ray transport in
  the gamma-ray sky}}.
\newblock \emph{\bibinfo{journal}{\jcap}} \textbf{\bibinfo{volume}{2017}},
  \bibinfo{pages}{019} (\bibinfo{year}{2017}).

\bibitem{Ahlers:2013xia}
\bibinfo{author}{Ahlers, M.} \& \bibinfo{author}{Murase, K.}
\newblock \bibinfo{title}{{Probing the Galactic Origin of the IceCube Excess
  with Gamma-Rays}}.
\newblock \emph{\bibinfo{journal}{Phys.Rev.}} \textbf{\bibinfo{volume}{D90}},
  \bibinfo{pages}{023010} (\bibinfo{year}{2014}).

\bibitem{2018PhRvD..98d3003L}
\bibinfo{author}{{Lipari}, P.} \& \bibinfo{author}{{Vernetto}, S.}
\newblock \bibinfo{title}{{Diffuse Galactic gamma-ray flux at very high
  energy}}.
\newblock \emph{\bibinfo{journal}{\prd}} \textbf{\bibinfo{volume}{98}},
  \bibinfo{pages}{043003} (\bibinfo{year}{2018}).

\bibitem{FangMurase21}
\bibinfo{author}{{Fang}, K.} \& \bibinfo{author}{{Murase}, K.}
\newblock \bibinfo{title}{{Multimessenger Implications of Sub-PeV Diffuse
  Galactic Gamma-Ray Emission}}.
\newblock \emph{\bibinfo{journal}{\apj}} \textbf{\bibinfo{volume}{919}},
  \bibinfo{pages}{93} (\bibinfo{year}{2021}).

\bibitem{2016ApJS..223...26A}
\bibinfo{author}{{Acero}, F.} \emph{et~al.}
\newblock \bibinfo{title}{{Development of the Model of Galactic Interstellar
  Emission for Standard Point-source Analysis of Fermi Large Area Telescope
  Data}}.
\newblock \emph{\bibinfo{journal}{\apjs}} \textbf{\bibinfo{volume}{223}},
  \bibinfo{pages}{26} (\bibinfo{year}{2016}).

\bibitem{4FGL}
\bibinfo{author}{{Abdollahi}, S.} \emph{et~al.}
\newblock \bibinfo{title}{{Fermi Large Area Telescope Fourth Source Catalog}}.
\newblock \emph{\bibinfo{journal}{\apjs}} \textbf{\bibinfo{volume}{247}},
  \bibinfo{pages}{33} (\bibinfo{year}{2020}).

\bibitem{2017JCAP...02..015E}
\bibinfo{author}{{Evoli}, C.} \emph{et~al.}
\newblock \bibinfo{title}{{Cosmic-ray propagation with DRAGON2: I. numerical
  solver and astrophysical ingredients}}.
\newblock \emph{\bibinfo{journal}{\jcap}} \textbf{\bibinfo{volume}{2017}},
  \bibinfo{pages}{015} (\bibinfo{year}{2017}).

\bibitem{2022arXiv220315759D}
\bibinfo{author}{{De La Torre Luque}, P.} \emph{et~al.}
\newblock \bibinfo{title}{{Galactic diffuse gamma rays meet the PeV frontier}}.
\newblock \emph{\bibinfo{journal}{\aap}} \textbf{\bibinfo{volume}{672}},
  \bibinfo{pages}{A58} (\bibinfo{year}{2023}).

\bibitem{2016JCAP...11..004P}
\bibinfo{author}{{Pagliaroli}, G.}, \bibinfo{author}{{Evoli}, C.} \&
  \bibinfo{author}{{Villante}, F.~L.}
\newblock \bibinfo{title}{{Expectations for high energy diffuse galactic
  neutrinos for different cosmic ray distributions}}.
\newblock \emph{\bibinfo{journal}{\jcap}} \textbf{\bibinfo{volume}{2016}},
  \bibinfo{pages}{004} (\bibinfo{year}{2016}).

\bibitem{2019JCAP...12..050C}
\bibinfo{author}{{Cataldo}, M.}, \bibinfo{author}{{Pagliaroli}, G.},
  \bibinfo{author}{{Vecchiotti}, V.} \& \bibinfo{author}{{Villante}, F.~L.}
\newblock \bibinfo{title}{{Probing galactic cosmic ray distribution with TeV
  gamma-ray sky}}.
\newblock \emph{\bibinfo{journal}{\jcap}} \textbf{\bibinfo{volume}{2019}},
  \bibinfo{pages}{050} (\bibinfo{year}{2019}).

\bibitem{2020A&A...641A...6P}
\bibinfo{author}{{Planck Collaboration}} \emph{et~al.}
\newblock \bibinfo{title}{{Planck 2018 results. VI. Cosmological parameters}}.
\newblock \emph{\bibinfo{journal}{\aap}} \textbf{\bibinfo{volume}{641}},
  \bibinfo{pages}{A6} (\bibinfo{year}{2020}).

\bibitem{1976ApJ...203..297S}
\bibinfo{author}{{Schechter}, P.}
\newblock \bibinfo{title}{{An analytic expression for the luminosity function
  for galaxies.}}
\newblock \emph{\bibinfo{journal}{\apj}} \textbf{\bibinfo{volume}{203}},
  \bibinfo{pages}{297--306} (\bibinfo{year}{1976}).

\bibitem{2006A&A...445..805F}
\bibinfo{author}{{Fasano}, G.} \emph{et~al.}
\newblock \bibinfo{title}{{WINGS: a WIde-field Nearby Galaxy-cluster Survey. I.
  Optical imaging}}.
\newblock \emph{\bibinfo{journal}{\aap}} \textbf{\bibinfo{volume}{445}},
  \bibinfo{pages}{805--817} (\bibinfo{year}{2006}).

\bibitem{2014ApJ...783...85T}
\bibinfo{author}{{Tomczak}, A.~R.} \emph{et~al.}
\newblock \bibinfo{title}{{Galaxy Stellar Mass Functions from ZFOURGE/CANDELS:
  An Excess of Low-mass Galaxies since z = 2 and the Rapid Buildup of Quiescent
  Galaxies}}.
\newblock \emph{\bibinfo{journal}{\apj}} \textbf{\bibinfo{volume}{783}},
  \bibinfo{pages}{85} (\bibinfo{year}{2014}).

\bibitem{2016ApJ...830...83C}
\bibinfo{author}{{Conselice}, C.~J.}, \bibinfo{author}{{Wilkinson}, A.},
  \bibinfo{author}{{Duncan}, K.} \& \bibinfo{author}{{Mortlock}, A.}
\newblock \bibinfo{title}{{The Evolution of Galaxy Number Density at z $<8$ and
  Its Implications}}.
\newblock \emph{\bibinfo{journal}{\apj}} \textbf{\bibinfo{volume}{830}},
  \bibinfo{pages}{83} (\bibinfo{year}{2016}).

\bibitem{2003ApJS..149..289B}
\bibinfo{author}{{Bell}, E.~F.}, \bibinfo{author}{{McIntosh}, D.~H.},
  \bibinfo{author}{{Katz}, N.} \& \bibinfo{author}{{Weinberg}, M.~D.}
\newblock \bibinfo{title}{{The Optical and Near-Infrared Properties of
  Galaxies. I. Luminosity and Stellar Mass Functions}}.
\newblock \emph{\bibinfo{journal}{\apjs}} \textbf{\bibinfo{volume}{149}},
  \bibinfo{pages}{289--312} (\bibinfo{year}{2003}).

\bibitem{2004MNRAS.349..146G}
\bibinfo{author}{{Gilfanov}, M.}
\newblock \bibinfo{title}{{Low-mass X-ray binaries as a stellar mass indicator
  for the host galaxy}}.
\newblock \emph{\bibinfo{journal}{\mnras}} \textbf{\bibinfo{volume}{349}},
  \bibinfo{pages}{146--168} (\bibinfo{year}{2004}).

\bibitem{2020A&A...641A.147K}
\bibinfo{author}{{Kornecki}, P.} \emph{et~al.}
\newblock \bibinfo{title}{{{\ensuremath{\gamma}}-ray/infrared luminosity
  correlation of star-forming galaxies}}.
\newblock \emph{\bibinfo{journal}{\aap}} \textbf{\bibinfo{volume}{641}},
  \bibinfo{pages}{A147} (\bibinfo{year}{2020}).

\bibitem{2020ApJ...889...32S}
\bibinfo{author}{{Suh}, H.} \emph{et~al.}
\newblock \bibinfo{title}{{No Significant Evolution of Relations between Black
  Hole Mass and Galaxy Total Stellar Mass Up to z {\ensuremath{\sim}} 2.5}}.
\newblock \emph{\bibinfo{journal}{\apj}} \textbf{\bibinfo{volume}{889}},
  \bibinfo{pages}{32} (\bibinfo{year}{2020}).

\bibitem{2006ApJ...651..142H}
\bibinfo{author}{{Hopkins}, A.~M.} \& \bibinfo{author}{{Beacom}, J.~F.}
\newblock \bibinfo{title}{{On the Normalization of the Cosmic Star Formation
  History}}.
\newblock \emph{\bibinfo{journal}{\apj}} \textbf{\bibinfo{volume}{651}},
  \bibinfo{pages}{142--154} (\bibinfo{year}{2006}).

\bibitem{HESE75yr}
\bibinfo{author}{{IceCube Collaboration}} \emph{et~al.}
\newblock \bibinfo{title}{{IceCube high-energy starting event sample:
  Description and flux characterization with 7.5 years of data}}.
\newblock \emph{\bibinfo{journal}{\prd}} \textbf{\bibinfo{volume}{104}},
  \bibinfo{pages}{022002} (\bibinfo{year}{2021}).

\bibitem{cascade6yr}
\bibinfo{author}{Aartsen, M.~G.} \emph{et~al.}
\newblock \bibinfo{title}{{Characteristics of the diffuse astrophysical
  electron and tau neutrino flux with six years of IceCube high energy cascade
  data}}.
\newblock \emph{\bibinfo{journal}{Phys. Rev. Lett.}}
  \textbf{\bibinfo{volume}{125}}, \bibinfo{pages}{121104}
  (\bibinfo{year}{2020}).

\bibitem{numu95yr}
\bibinfo{author}{Abbasi, R.} \emph{et~al.}
\newblock \bibinfo{title}{{Improved Characterization of the Astrophysical
  Muon\textendash{}neutrino Flux with 9.5 Years of IceCube Data}}.
\newblock \emph{\bibinfo{journal}{Astrophys. J.}}
  \textbf{\bibinfo{volume}{928}}, \bibinfo{pages}{50} (\bibinfo{year}{2022}).

\bibitem{1998PhRvD..59b3002W}
\bibinfo{author}{{Waxman}, E.} \& \bibinfo{author}{{Bahcall}, J.}
\newblock \bibinfo{title}{{High energy neutrinos from astrophysical sources: An
  upper bound}}.
\newblock \emph{\bibinfo{journal}{\prd}} \textbf{\bibinfo{volume}{59}},
  \bibinfo{pages}{023002} (\bibinfo{year}{1998}).

\bibitem{FermiIGRB}
\bibinfo{author}{{Ackermann}, M.} \emph{et~al.}
\newblock \bibinfo{title}{{The Spectrum of Isotropic Diffuse Gamma-Ray Emission
  between 100 MeV and 820 GeV}}.
\newblock \emph{\bibinfo{journal}{\apj}} \textbf{\bibinfo{volume}{799}},
  \bibinfo{pages}{86} (\bibinfo{year}{2015}).

\bibitem{2007ARNPS..57..285S}
\bibinfo{author}{{Strong}, A.~W.}, \bibinfo{author}{{Moskalenko}, I.~V.} \&
  \bibinfo{author}{{Ptuskin}, V.~S.}
\newblock \bibinfo{title}{{Cosmic-Ray Propagation and Interactions in the
  Galaxy}}.
\newblock \emph{\bibinfo{journal}{Annual Review of Nuclear and Particle
  Science}} \textbf{\bibinfo{volume}{57}}, \bibinfo{pages}{285--327}
  (\bibinfo{year}{2007}).

\bibitem{Fermi-LAT:2014sfa}
\bibinfo{author}{Ackermann, M.} \emph{et~al.}
\newblock \bibinfo{title}{{The Spectrum and Morphology of the $Fermi$
  Bubbles}}.
\newblock \emph{\bibinfo{journal}{Astrophys. J.}}
  \textbf{\bibinfo{volume}{793}}, \bibinfo{pages}{64} (\bibinfo{year}{2014}).

\bibitem{Predehl:2020kyq}
\bibinfo{author}{Predehl, P.} \emph{et~al.}
\newblock \bibinfo{title}{{Detection of large-scale X-ray bubbles in the Milky
  Way halo}}.
\newblock \emph{\bibinfo{journal}{Nature}} \textbf{\bibinfo{volume}{588}},
  \bibinfo{pages}{227--231} (\bibinfo{year}{2020}).

\bibitem{Yang:2022jck}
\bibinfo{author}{Yang, H. Y.~K.}, \bibinfo{author}{Ruszkowski, M.} \&
  \bibinfo{author}{Zweibel, E.~G.}
\newblock \bibinfo{title}{{Fermi and eROSITA bubbles as relics of the past
  activity of the Galaxy\textquoteright{}s central black hole}}.
\newblock \emph{\bibinfo{journal}{Nature Astron.}}
  \textbf{\bibinfo{volume}{6}}, \bibinfo{pages}{584--591}
  (\bibinfo{year}{2022}).

\bibitem{2015ApJ...799L...7F}
\bibinfo{author}{{Fox}, A.~J.} \emph{et~al.}
\newblock \bibinfo{title}{{Probing the Fermi Bubbles in Ultraviolet Absorption:
  A Spectroscopic Signature of the Milky Way's Biconical Nuclear Outflow}}.
\newblock \emph{\bibinfo{journal}{\apjl}} \textbf{\bibinfo{volume}{799}},
  \bibinfo{pages}{L7} (\bibinfo{year}{2015}).

\bibitem{2016ApJ...829....9M}
\bibinfo{author}{{Miller}, M.~J.} \& \bibinfo{author}{{Bregman}, J.~N.}
\newblock \bibinfo{title}{{The Interaction of the Fermi Bubbles with the Milky
  Way{\textquoteright}s Hot Gas Halo}}.
\newblock \emph{\bibinfo{journal}{\apj}} \textbf{\bibinfo{volume}{829}},
  \bibinfo{pages}{9} (\bibinfo{year}{2016}).

\bibitem{2017ApJ...834..191B}
\bibinfo{author}{{Bordoloi}, R.} \emph{et~al.}
\newblock \bibinfo{title}{{Mapping the Nuclear Outflow of the Milky Way:
  Studying the Kinematics and Spatial Extent of the Northern Fermi Bubble}}.
\newblock \emph{\bibinfo{journal}{\apj}} \textbf{\bibinfo{volume}{834}},
  \bibinfo{pages}{191} (\bibinfo{year}{2017}).

\bibitem{2019ApJ...886...45B}
\bibinfo{author}{{Bland-Hawthorn}, J.} \emph{et~al.}
\newblock \bibinfo{title}{{The Large-scale Ionization Cones in the Galaxy}}.
\newblock \emph{\bibinfo{journal}{\apj}} \textbf{\bibinfo{volume}{886}},
  \bibinfo{pages}{45} (\bibinfo{year}{2019}).

\bibitem{2016PhRvL.116g1101M}
\bibinfo{author}{{Murase}, K.}, \bibinfo{author}{{Guetta}, D.} \&
  \bibinfo{author}{{Ahlers}, M.}
\newblock \bibinfo{title}{{Hidden Cosmic-Ray Accelerators as an Origin of
  TeV-PeV Cosmic Neutrinos}}.
\newblock \emph{\bibinfo{journal}{\prl}} \textbf{\bibinfo{volume}{116}},
  \bibinfo{pages}{071101} (\bibinfo{year}{2016}).

\bibitem{2022ApJ...933..190F}
\bibinfo{author}{{Fang}, K.}, \bibinfo{author}{{Gallagher}, J.~S.} \&
  \bibinfo{author}{{Halzen}, F.}
\newblock \bibinfo{title}{{The TeV Diffuse Cosmic Neutrino Spectrum and the
  Nature of Astrophysical Neutrino Sources}}.
\newblock \emph{\bibinfo{journal}{\apj}} \textbf{\bibinfo{volume}{933}},
  \bibinfo{pages}{190} (\bibinfo{year}{2022}).

\bibitem{2022Sci...378..538I}
\bibinfo{author}{{IceCube Collaboration}} \emph{et~al.}
\newblock \bibinfo{title}{{Evidence for neutrino emission from the nearby
  active galaxy NGC 1068}}.
\newblock \emph{\bibinfo{journal}{Science}} \textbf{\bibinfo{volume}{378}},
  \bibinfo{pages}{538--543} (\bibinfo{year}{2022}).

\end{thebibliography}



\section*{Supplementary material}

# Supplemental Material

**Contents**

# 1 Model of the Cosmic Ray Densities

As Galactic cosmic rays are mostly light particles, we model the hadronic cosmic-ray population as protons and helium nulcei. The measurements of the cosmic-ray proton and helium spectra by various experiments are shown in Supplementary Figure 1. To follow the spectral shape, in particular the breaks around the rigidity 300 GV, the cosmic-ray spectrum is modeled with a modified power law function [1] that cuts off at low and high energies,

$$\Phi(R) = N_0 \left(\frac{R}{R_{\text{ref}}}\right)^\gamma \left[1 + \left(\frac{R}{R_0}\right)^{\Delta\gamma/s}\right]^s \exp\left[-\left(\frac{R_l}{R}\right)^{p_l}\right] \exp\left[-\left(\frac{R}{R_h}\right)^{p_h}\right]. \quad (S1)$$

We adopt $R_{\text{ref}} \equiv 45$ GV, and fix the values of $\gamma$, $R_0$, and $s$ to the best-fit parameters in [1,2], which were found by fitting the double power law function to the AMS-02 proton (helium) data between 45 GV and 1.8 (3) TV. We fix $p_h = 0.5$ since the model is poorly constrained by the data above the knee. The remaining parameters, namely, the lower cutoff rigidity $R_l$ and index $p_l$, the higher cutoff rigidity $R_h$, and the normalization $N_0$ are obtained by fitting equation S1 to the observational data listed in the caption of Supplementary Figure 1. Above $\sim 500$ TeV, the fit of the proton spectrum only used the IceTop data while the fit of helium spectrum used all data. The model parameters are summarized in Table 1.

The fits to proton and helium data yielded $\chi^2/d.o.f. \approx 0.3$ and 0.4, respectively. The relatively low reduced $\chi^2$ values are due to an over-estimation of the errors, since the errors in a lot of the measurements are dominated by systematic uncertainties that are correlated across energy bins. The best-fit spectra are used as cosmic-ray spectral shapes in the Galaxy to calculate the secondary production. The fits do not disentangle the systematic and statistical errors of the data nor account

2

Supplementary Table 1: Parameters in equation S1 obtained by fitting to direct and indirect cosmic-ray observations. Starred parameters are fixed in the fits.

| Composition | $N_0$ | $R^*_{\text{ref}}$ | $R^*_0$ | $\Delta\gamma$ | $s^*$ | $R_l$ | $p_l$ | $R_h$ | $p^*_h$ |
|---|---|---|---|---|---|---|---|---|---|
| | [GV$^{-1}$cm$^{-2}$s$^{-1}$sr$^{-1}$] | [GV] | [GV] | | | [GV] | | [PV] | |
| proton | $4.3 \times 10^{-5}$ | 45 | 336 | 0.24 | 0.024 | 3.8 | 0.86 | 10.2 | 0.5 |
| helium | $3.1 \times 10^{-6}$ | 45 | 245 | 0.28 | 0.027 | 4.2 | 0.60 | 0.64 | 0.5 |

for the energy calibrations across experiments. The lower cutoff $R_l$ is implemented to mimic the solar modulation effect [12], though the solar modulation is irrelevant for the energies considered in this work. For these reasons we caution that the best-fit spectra in this work should not be used to study the primary cosmic rays.

The cosmic-ray density in our model is computed using the public simulation framework CRPropa 3.1 [13]. Stochastic differential equations are solved to describe particle propagation in the Galactic magnetic field described with JF12 model, including both the regular and turbulent components [14]. The trajectories for 6 million pseudo particles with an injection rigidity of 100 TV are tracked until the maximum integration time of 10 Myr or the escaping from the 3D simulation volume at $\vec{x}_{\text{max}} = 20$ kpc. Our simulation uses adaptive steps between $0.1\,\text{pc}/c$ and $1\,\text{kpc}/c$ and a relative precision $\eta = 10^{-4}$. The sources are assumed to follow the distribution of isolated radio pulsars [15,16] which trace the sprial structure of the Galaxy. The spiral arm source density model

3

produces a higher cosmic-ray energy density than a smooth disk when the propagated intensities for each are similarly normalized to the cosmic-ray data. A face-on view of the cosmic-ray density on the plane ($b = 0°$) is shown in Supplementary Figure 2.

Our simulation adopted the cosmic-ray electron and positron model of Ref. [17], which was computed using the `DRAGON` code [18,19] under the assumption of a homogeneous and isotropic diffusion coefficient, and tabulated as the `Dragon2D` class in the `HERMES` code. Since the contribution of leptons to the GDE-$\gamma$ is sub-dominant above $\sim 1$ GeV, the details with the modeling of leptons barely impact our results.

Our model assumed that the spatial and spectral components of the nucleon flux are independent. This assumption could be overly simplified and inconsistent with *Fermi*-LAT observations [20]. Variations of the cosmic-ray spectra across the Galaxy also affect the modeling of the GDE-$\nu$ flux [21].

## 2 Secondary Production

The emissivity and intensity of secondary production is computed using the `HERMES` code [27]. The calculation accounts for the cosmic-ray interaction with neutral (HI) and molecular hydrogen (H$_2$) traced by the CO emission using the proton-proton interaction cross section from Refs. [28]. As in reference [27], we assume that the ISM gas is composed of hydrogen and helium nuclei with a uniform abundance ratio $n_{\text{He}}/n_{\text{H}} = 0.1$. The secondary emission is the sum of the products from proton and helium cosmic rays colliding with hydrogen and helium gas. To calculate both the

4

emissivity of the entire Galaxy and the intensity observed at the solar neighborhood, our calculation uses the 3D gas density profiles instead of the column density as in the original `HERMES` code. We adopt the gas density profiles from Refs. [29,30]. Above 3 kpc, we use the analytical model of Ref. [31] to describe the HI density distribution. We compared the secondary spectra calculated using the gas density profile directly and the column density skymaps of Galactocentric "rings". We find that the fluxes from the interaction with the HI gas are similar between the two. The fluxes from the interaction with the $H_2$ gas are comparable when the CO-$H_2$ conversion factor, $X_{\rm CO}$, in the latter approach is set to $\sim 2$ in units of $10^{20}\,{\rm cm}^{-2}{\rm K}^{-1}/({\rm km\,s}^{-1})$. Our gas model does not include ionized hydrogen (HII, e.g., ref [32]) and dark neutral medium gas [33], though contribution from these gas components is minor. Our simulation used a linear 3D grid with $0.25$ kpc spacing in x and y directions from -25 kpc to 25 kpc and in z direction from -5 kpc to 5 kpc.

Supplementary Figure 3 further decomposes the total emission into contributions from nuclei, including their interaction with HI gas and molecular hydrogen traced by CO, and electrons, including their inverse Compton and synchrotron radiation. In *region A*, our model explains the IceCube and the Tibet AS$\gamma$ GDE observations but overproduces the LHAASO-KM2A measurement, which is $\sim$2-3 times lower than the Tibet AS$\gamma$ measurement. The difference in multi-messenger observations could be attributed to unresolved sources. The source contribution, on the other hand, should not alter the conclusion that the hadronic emission dominates the Galactic plane emission above $\sim 30$ TeV given the consistency in the IceCube and Tibet AS$\gamma$ fluxes.

The longitudinal and latitudinal distributions of our GDE-$\gamma$ model are roughly consistent

5

with *Fermi*-LAT observations above 100 GeV. Nevertheless, such a simplified model is not tailored to fit the data, which is known to be explained by multiple emission components [20].

The light blue bands in Supplementary Figure 3 correspond to the average intensity of the two Tibet regions in the $\pi^0$ template but scaled by the full-sky normalization found by the IceCube analysis. They are not measurements using neutrino events in the corresponding sky regions. Comparing to them, our model produces more flux in *region A* and less flux in *region B*, suggesting that our model has a different spatial distribution from the $\pi^0$ template.

Supplementary Figure 4 compares our model with various observations of the Galactic plane. Our model agrees with the measurements using the *Fermi* $\pi^0$ and CRINGE models but overproduces those using the KRA models. We note that the model presented in this work is designed to explain the IceCube $\pi^0$ flux as diffuse emission. When changing our source density distribution from a spiral arm model to a smooth disk model, the prediction of the secondary flux would be lowered by a factor of $\sim 2$ and comparable to the $\pi^0$ and KRA models at $\sim 100$ GeV.

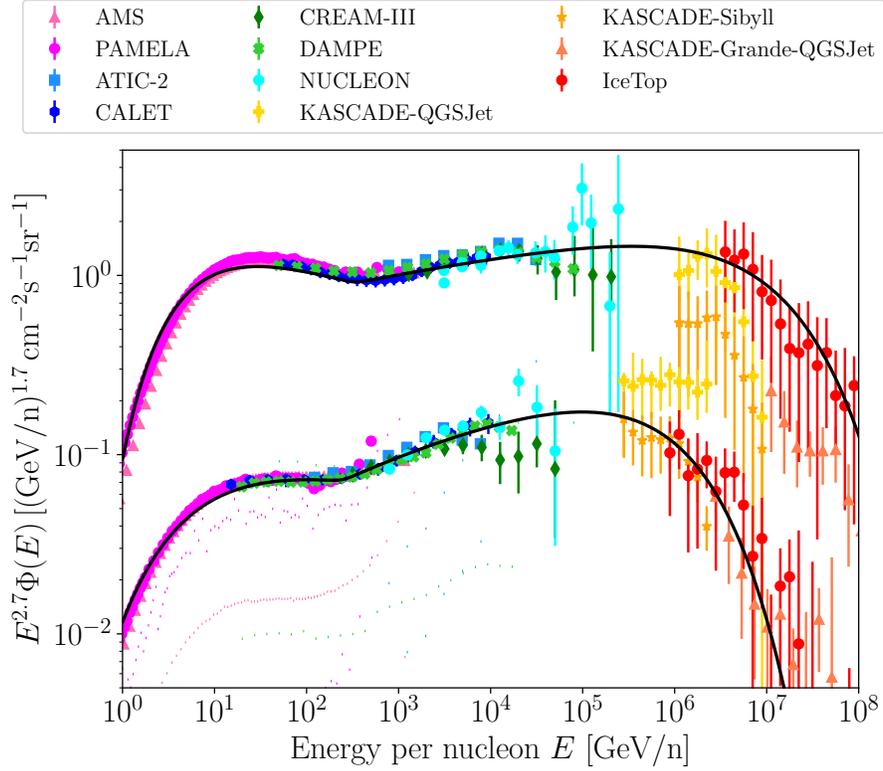

Supplementary Figure 1: **Cosmic-ray proton and helium fluxes at the Earth.** The top and bottom curves indicate proton and helium fluxes, respectively. The data points are from measurements by AMS-02 [1,2], PAMELA [3], ATIC [4], CALET [5,6], CREAM-III [7], DAMPE [8], NUCLEON [9], KASCADE [10], and IceTop [11] experiments. The errors indicate the quadratic sums of statistical and systematic uncertainties of the measurements. The black curves are model fits to the data points.

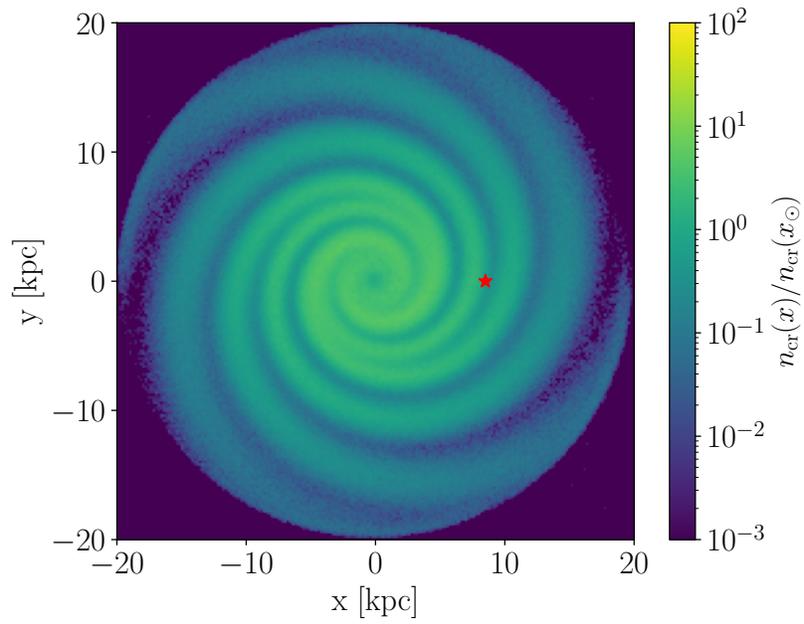

Supplementary Figure 2: **Face-on view of the Galactic cosmic-ray density at 100 TeV at $b = 0°$.** The color scale indicates the ratio of the number density of cosmic rays at a given position $\vec{x}$ to that of the solar neighborhood $\vec{x}_\odot$. The red star marks the location of the sun.

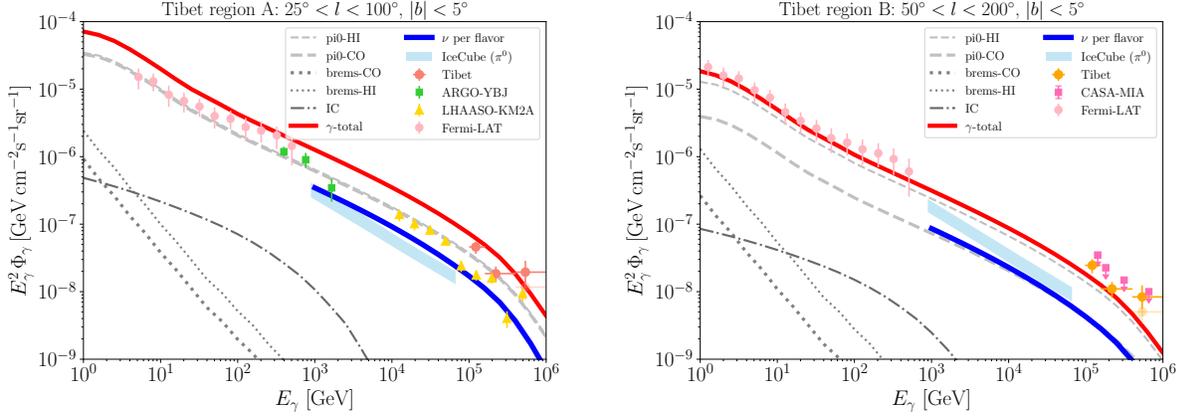

Supplementary Figure 3: **Intensities of Galactic diffuse emission in $\gamma$-ray and neutrino from two sky regions,** *region A* **(left):** $25° < l < 100°$, $|b| < 5°$, **and** *region B* **(right):** $50° < l < 200°$, $|b| < 5°$. The data points refer to observations of various $\gamma$-ray experiments, including Tibet AS$\gamma$ [22] (orange data points; the fainter data points at 500 TeV indicate the residual intensity after removing events near the Cygnus Cocoon), LHAASO-KM2A [23] (yellow triangle markers), *Fermi*-LAT [24] (pink circular markers indicating the data of the corresponding region), ARGO-YBJ [25] (green circular markers in *region A*), and CASA-MIA [26] (magenta square markers in *region B*). The error bars show $1\,\sigma$ statistical errors. The $\gamma$-ray contribution (red) is decomposed into $\pi^0$ decay (grey dashed), inverse Compton (grey dash-dotted) and bremsstrahlung emission (grey dotted), with thick and thin curves indicating products from interaction with molecular ($H_2$) and neutral (HI) gas, respectively. The blue curves indicate the corresponding per-flavor neutrino intensities. The $\gamma$-ray and neutrino spectra presented in this figure are from a simple model where the spatial and spectral components of the nucleon flux are assumed to be independent.

9

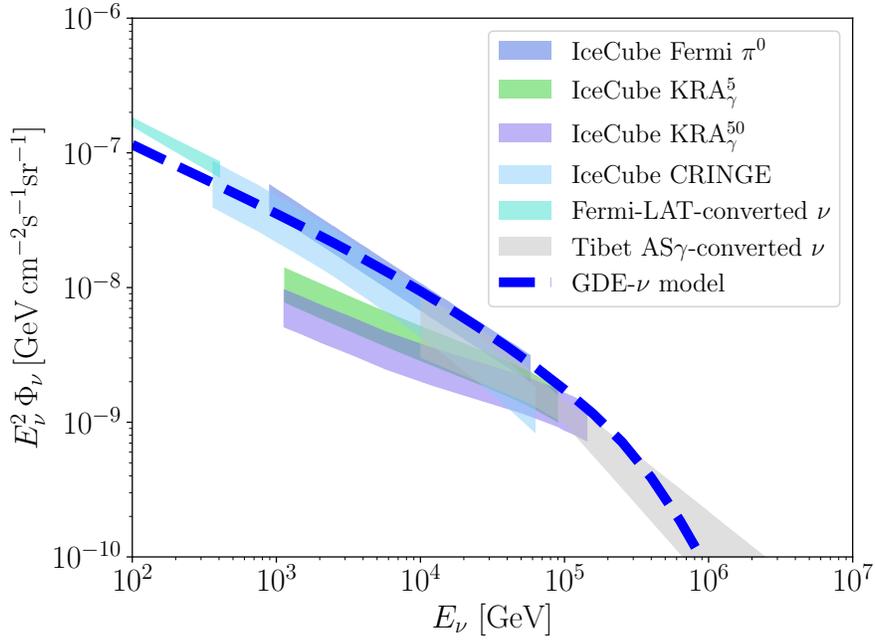

Supplementary Figure 4: **Comparison of our Galactic neutrino diffuse emission model with the observed and derived flux of the Galactic Plane.** The blue dashed curve indicates the model. The IceCube observations include measurements using cascade events and the *Fermi* $\pi^0$ model (dark blue band), KRA models (green and purple bands) [34] and using track events and the CRINGE model (light blue band) [35,36]. Also shown are the neutrino flux derived from the GDE-$\gamma$ measured by Tibet AS$\gamma$ [22,37] (silver band) and the Galactic interstellar emission model of *Fermi*-LAT [38] (turquoise band). All curves except the CRINGE band are $4\pi$-averaged. The IceCube CRINGE flux is based on an analysis with track data from the northern hemisphere and is averaged over the corresponding sensitive sky region.

1. Aguilar, M. *et al.* Precision measurement of the proton flux in primary cosmic rays from rigidity 1 gv to 1.8 tv with the alpha magnetic spectrometer on the international space station. *Phys. Rev. Lett.* **114**, 171103 (2015). URL https://link.aps.org/doi/10.1103/PhysRevLett.114.171103.

2. Aguilar, M. *et al.* Precision measurement of the helium flux in primary cosmic rays of rigidities 1.9 gv to 3 tv with the alpha magnetic spectrometer on the international space station. *Phys. Rev. Lett.* **115**, 211101 (2015). URL https://link.aps.org/doi/10.1103/PhysRevLett.115.211101.

3. Adriani, O. *et al.* PAMELA Measurements of Cosmic-Ray Proton and Helium Spectra. *Science* **332**, 69 (2011). 1103.4055.

4. Panov, A. D. *et al.* Energy spectra of abundant nuclei of primary cosmic rays from the data of ATIC-2 experiment: Final results. *Bulletin of the Russian Academy of Sciences, Physics* **73**, 564–567 (2009). 1101.3246.

5. Calet Collaboration *et al.* Direct Measurement of the Cosmic-Ray Proton Spectrum from 50 GeV to 10 TeV with the Calorimetric Electron Telescope on the International Space Station. *Phys. Rev. Lett.* **122**, 181102 (2019). 1905.04229.

6. Brogi, P. & Kobayashi, K. Measurement of the energy spectrum of cosmic-ray helium with CALET on the International Space Station. In *Proceedings of 37th International Cosmic Ray Conference — PoS(ICRC2021)*, vol. 395, 101 (2021).

7. Yoon, Y. S. *et al.* Proton and Helium Spectra from the CREAM-III Flight. *ApJ.* **839**, 5 (2017). 1704.02512.

8. An, Q. *et al.* Measurement of the cosmic ray proton spectrum from 40 GeV to 100 TeV with the DAMPE satellite. *Science Advances* **5**, eaax3793 (2019). 1909.12860.

9. Atkin, E. *et al.* First results of the cosmic ray NUCLEON experiment. *JCAP* **07**, 020 (2017). 1702.02352.

10. Apel, W. D. *et al.* KASCADE-Grande measurements of energy spectra for elemental groups of cosmic rays. *Astroparticle Physics* **47**, 54–66 (2013).

11. Aartsen, M. G. *et al.* Cosmic ray spectrum and composition from PeV to EeV using 3 years of data from IceTop and IceCube. *Phys. Rev. D* **100**, 082002 (2019). 1906.04317.

12. Gleeson, L. J. & Axford, W. I. Solar Modulation of Galactic Cosmic Rays. *ApJ.* **154**, 1011 (1968).

13. Merten, L., Becker Tjus, J., Fichtner, H., Eichmann, B. & Sigl, G. CRPropa 3.1—a low energy extension based on stochastic differential equations. *JCAP* **2017**, 046 (2017). 1704.07484.

14. Jansson, R. & Farrar, G. R. A NEW MODEL OF THE GALACTIC MAGNETIC FIELD. *The Astrophysical Journal* **757**, 14 (2012). URL https://doi.org/10.1088/0004-637x/757/1/14.

15. Faucher-Giguère, C.-A. & Kaspi, V. M. Birth and Evolution of Isolated Radio Pulsars. *ApJ.* **643**, 332–355 (2006). astro-ph/0512585.

16. Blasi, P. & Amato, E. Diffusive propagation of cosmic rays from supernova remnants in the Galaxy. I: spectrum and chemical composition. *JCAP* **2012**, 010 (2012). `1105.4521`.

17. Fornieri, O., Gaggero, D. & Grasso, D. Features in cosmic-ray lepton data unveil the properties of nearby cosmic accelerators. *JCAP* **2020**, 009 (2020). `1907.03696`.

18. Evoli, C. *et al.* Cosmic-ray propagation with DRAGON2: I. numerical solver and astrophysical ingredients. *JCAP* **2017**, 015 (2017). `1607.07886`.

19. Evoli, C. *et al.* Cosmic-ray propagation with DRAGON2: II. Nuclear interactions with the interstellar gas. *JCAP* **2018**, 006 (2018). `1711.09616`.

20. Acero, F. *et al.* Development of the Model of Galactic Interstellar Emission for Standard Point-source Analysis of Fermi Large Area Telescope Data. *The Astrophysical Journal Supplement Series* **223**, 26 (2016). `1602.07246`.

21. Gaggero, D., Grasso, D., Marinelli, A., Urbano, A. & Valli, M. The Gamma-Ray and Neutrino Sky: A Consistent Picture of Fermi-LAT, Milagro, and IceCube Results. *ApJ Letters* **815**, L25 (2015). `1504.00227`.

22. Tibet AS$\gamma$ Collaboration *et al.* First Detection of sub-PeV Diffuse Gamma Rays from the Galactic Disk: Evidence for Ubiquitous Galactic Cosmic Rays beyond PeV Energies. *Phys. Rev. Lett.* **126**, 141101 (2021). `2104.05181`.

23. Zhao, S., Zhang, R., Zhang, Y. & Yuan, Q. Measurement of the diffuse gamma-ray emission from Galactic plane with LHAASO-KM2A. In *Proceedings of 37th International Cosmic Ray Conference — PoS(ICRC2021)*, vol. 395, 859 (2021).

24. De La Torre Luque, P. *et al.* Galactic diffuse gamma rays meet the PeV frontier. *A&A* **672**, A58 (2023). 2203.15759.

25. Bartoli, B. *et al.* Study of the Diffuse Gamma-ray Emission From the Galactic Plane With ARGO-YBJ. *Astrophys. J.* **806**, 20 (2015). 1507.06758.

26. Borione, A. *et al.* Constraints on gamma-ray emission from the galactic plane at 300 TeV. *The Astrophysical Journal* **493**, 175–179 (1998). URL https://doi.org/10.1086/305096.

27. Dundovic, A., Evoli, C., Gaggero, D. & Grasso, D. Simulating the Galactic multi-messenger emissions with HERMES. *A&A* **653**, A18 (2021). 2105.13165.

28. Kelner, S. R., Aharonian, F. A. & Bugayov, V. V. Energy spectra of gamma rays, electrons, and neutrinos produced at proton-proton interactions in the very high energy regime. *Physical Review D* **74**, 034018 (2006). astro-ph/0606058.

29. Nakanishi, H. & Sofue, Y. Three-Dimensional Distribution of the ISM in the Milky Way Galaxy: I. The H I Disk. *Publications of the ASJ* **55**, 191–202 (2003). astro-ph/0304338.

30. Nakanishi, H. & Sofue, Y. Three-Dimensional Distribution of the ISM in the Milky Way Galaxy: II. The Molecular Gas Disk. *Publications of the ASJ* **58**, 847–860 (2006). astro-ph/0610769.

31. Ferrière, K. Global Model of the Interstellar Medium in Our Galaxy with New Constraints on the Hot Gas Component. *ApJ.* **497**, 759–776 (1998).

32. Yao, J. M., Manchester, R. N. & Wang, N. A New Electron-density Model for Estimation of Pulsar and FRB Distances. *ApJ.* **835**, 29 (2017). `1610.09448`.

33. Planck Collaboration *et al.* Planck intermediate results. XXVIII. Interstellar gas and dust in the Chamaeleon clouds as seen by Fermi LAT and Planck. *A&A* **582**, A31 (2015). `1409.3268`.

34. IceCube Collaboration. Observation of high-energy neutrinos from the galactic plane. *Science* **380**, 1338–1343 (2023). URL `https://www.science.org/doi/abs/10.1126/science.adc9818`.

35. Schwefer, G., Mertsch, P. & Wiebusch, C. Diffuse Emission of Galactic High-energy Neutrinos from a Global Fit of Cosmic Rays. *ApJ.* **949**, 16 (2023). `2211.15607`.

36. Fuerst, P. M. *et al.* Galactic and Extragalactic Analysis of the Astrophysical Muon Neutrino Flux with 12.3 years of IceCube Track Data. *PoS* **ICRC2023**, 1046 (2023).

37. Fang, K. & Murase, K. Multimessenger Implications of Sub-PeV Diffuse Galactic Gamma-Ray Emission. *ApJ.* **919**, 93 (2021). `2104.09491`.

38. Abdollahi, S. *et al.* Fermi Large Area Telescope Fourth Source Catalog. *The Astrophysical Journal Supplement Series* **247**, 33 (2020). `1902.10045`.



\end{document}